\documentstyle[10pt]{article}
\normalbaselines
\begin{document}

\baselineskip=12pt

\bibliographystyle{unsrt}
\vbox{\vspace{6mm}}

\begin{center}
{\large\bf Stochastic control of quantum dynamics \\
for a single trapped system}
\end{center}

\begin{center}
{Stefano Mancini}\footnote{email: mancini@mi.infn.it}
\end{center}

\begin{center}
{\it INFM, Dipartimento di Fisica,
Universit\`a di Milano, \\
Via Celoria 16, I-20133 Milano, Italy}
\end{center}

\begin{abstract}
A stochastic control of the vibrational motion
for a single trapped ion/atom is proposed.
It is based on the possibility to continously monitor 
the motion through a light field meter.
The output from the measurement process should be then used 
to modify the system's dynamics.
\end{abstract}

\section{Introduction}

In recent years there has been an increasing interest on trapping 
phenomena and related cooling techniques \cite{1}.
A few years ago, it has been shown that a single ion/atom can be trapped 
and cooled down near to its zero-point vibrational energy state \cite{2}. 
On the other hand,
the continuous measurement of quantum systems is
particularly relevant at this time because experimental technology is now
at the point where {\em individual} quantum systems can be monitored
continuously \cite{3}. With these developments it should be
possible in the near future to control quantum systems in real time by
using the results of the measurement in the process of continuous
feedback.
The possibility to control trapped particles, 
indeed, gave rise to new models
in quantum computation \cite{4}. In these systems the 
two lowest vibrational states
are used for logical operations among quantum bits and the 
dominant source of decoherence is 
the heating of the vibrational motion \cite{5}.
Here, a way to control the position of a trapped ion/atom is studied.
It results the possibility of a motional phase space uncertainty 
contraction corresponding to localise the particle
even below the quantum limit.

\section{The Measurement Model}

Let us consider one vibrational mode for a trapped system
together with a meter mode. The Hamiltonian 
would be of the following form
\newpage
\begin{equation}\label{Hgen}
H=H_{m}
+\hbar\nu a^{\dag}a+\hbar\chi{\cal O}X\,,
\end{equation}
where $H_m$ is the free Hamiltonian of the meter,
$\nu$ being the oscillation frequency of the ion/atom in the trap,
and $a$, $a^{\dag}$ the lowering and rising operators for 
the vibrational states of the trap.
In the last term of Eq.(\ref{Hgen}),
$\chi$ is the coupling constant,
$X=(a+a^{\dag})/2$ is the dimensionless position operator, and
${\cal O}$ can be any meter operator obtained as linear combination 
of the $H_{m}$'s eigenoperators ${\cal O}^{\pm}$ (but, for simplicity,
in the following it is assumed as the sum of them).
The practical implementation of such type of Hamiltonian will 
be discussed later on.

The evolution 
equation for the whole density operator $D$ is assumed to be
\begin{equation}
\dot{D}={\cal L}D -i\chi\left[{\cal O}X,D\right]
+\frac{\kappa}{2}
\left(2{\cal O}^{-}D{\cal O}^{+}-{\cal O}^+{\cal O}^{-}D
-D{\cal O}^{+}{\cal O}^{-}\right)\;,
\label{evoluzero}
\end{equation}
where ${\cal L}$ describes the damped dynamics of the 
vibrational mode.
Furthermore, the meter mode is considered heavily damped, 
so that the decay rate 
$\kappa$ is very large, ($\kappa \gg \chi$), and the meter
will almost always  be 
in the lower eigenstate of $H_m$. 
This allows the adiabatic elimination of the
meter mode 
by means of a perturbative calculation in the small 
parameter $\chi /\kappa$, obtaining (see also Ref. \cite{6})             
\begin{equation}\label{Dofrho}
D=\left[\rho\otimes\Pi\right]
+\left(\frac{\chi}{\kappa}\right)\left[
i\rho X\otimes\Pi{\cal O}^{-}+h.c.\right]
+\left(\frac{\chi}{\kappa}\right)^2
\left[\dots\right]
+\ldots\,,
\end{equation}
where $\rho={\rm Tr}_{m}\,D$ is the reduced density matrix for 
the vibrational motion, and $\Pi$ is the projector on the 
lower eigenstate of $H_m$.

Let us now soppose to be able to perform an  
homodyne-like measurement on the meter mode, i.e. to measure the
observable ${\cal O}_{\varphi}=({\cal O}^-e^{-i\varphi}
+{\cal O}^+e^{i\varphi})/2$.
Then, the current at readout apparatus
will be \cite{7}
\begin{equation}\label{photoc}
I(t)=2\eta\kappa\langle{\cal O}_{\varphi}(t)\rangle_c
+\sqrt{\eta\kappa}\,\xi(t)\,,
\end{equation}
where $\eta$ is the
efficiency of the measurement process.
The subscript
$c$ in Eq.~(\ref{photoc}) 
denotes the fact that the average is performed on the state 
conditioned on
the results of the previous measurements and
$\xi(t)$ is a Gaussian white noise \cite{7}.
Therefore, due to the interaction in Eq.(1), one gets 
information 
on $X$ by observing the quantity ${\cal O}_{\varphi}$. 
From Eq.(\ref{Dofrho}), the relationship between the conditioned 
mean values results 
$\langle{\cal O}_{\varphi}\rangle_c
=(\chi/\kappa)\,\langle X\rangle_c\,\sin\varphi$.

However, the continuous monitoring of the meter mode
modifies the time evolution of the whole system. In fact, the 
state conditioned on the result of
measurement, described by the conditioned density matrix 
$D_c$, evolves
according to the following stochastic differential equation 
(considered in the Ito sense)
\begin{eqnarray}\label{Dceq}
{\dot D}_c&=&{\cal L}D_c-i\chi
\left[{\cal O} X,D_c\right]
+\frac{\kappa}{2}\left(2{\cal O}^{-}D_c{\cal O}^{+}-{\cal O}^{+}{\cal O}^{-}
D_c-D_c{\cal O}^{+}{\cal O}^{-}\right)
\nonumber\\
&+&\sqrt{\eta\kappa}\,\xi(t)\left(
e^{-i\varphi}{\cal O}^{-}D_c+e^{i\varphi}D_c{\cal O}^{+}
-2\langle{\cal O}_{\varphi}\rangle_c\,D_c\right)\,.
\end{eqnarray}
If we adopt the perturbative expression (\ref{Dofrho}) for 
$D_c$ in (\ref{Dceq}) and perform
the trace over the meter mode, we get an equation for the 
reduced density matrix $\rho_c$
conditioned to the result of the measurement of the observable 
${\cal O}_{\varphi}$
\begin{equation}\label{rhoceq}
{\dot\rho}_c={\cal L}\rho_c-\frac{\chi^2}{2\kappa}
\left[X,\left[X,\rho_c\right]\right]
+\sqrt{\eta\chi^2/\kappa}\,\xi(t)
\left(ie^{i\varphi}\rho_cX-ie^{-i\varphi}X\rho_c
+2\sin\varphi\langle X\rangle_c\,\rho_c\right)\,.
\end{equation}

\section{Applying the feedback loop}

Let us now consider the application of a feedback loop 
to control the dynamics of the vibrational mode. 
The continous feedback theory proposed by Wiseman and Milburn 
\cite{8} is then invoked.
One has to take part 
of the stochastic output current $I(t)$, 
and feed it back to the vibrational dynamics 
in order to modify this. 
To be more specific, the presence of feedback modifies the 
evolution of the conditioned state
$\rho_c(t)$. It is reasonable to assume that the feedback 
effect can be described by an additional
term in the master equation, linear in the photocurrent 
$I(t)$, i.e. \cite{8}
\begin{equation}\label{rhofb}
\left[{\dot\rho}_c(t)\right]_{fb}=\frac{I(t-\tau)}{\eta\chi}\,
{\cal K}\rho_c(t)\,,
\end{equation}
where $\tau$ is the time delay in the feedback loop
and ${\cal K}$ 
is a Liouville superoperator describing the way 
in which the feedback signal acts 
on the system of interest.  

The feedback term (\ref{rhofb}) has to be considered in the 
Stratonovich sense, since Eq.
(\ref{rhofb}) is introduced as limit of a real process, then 
it should be transformed in the Ito
sense and added to the evolution equation (\ref{rhoceq}). 
A successive average over the white noise
$\xi(t)$ yields, in the limiting case of $\tau\to 0$,
the master equation  
\begin{equation}
\dot{\rho }={\cal L}\rho -\frac{\chi^2}{2\kappa}\left[X,\left[X 
,\rho \right]\right]+{\cal K}\left(ie^{i\varphi }\rho X-
ie^{-i\varphi} X\rho\right)
+\frac{{\cal K}^{2}}{2\eta\chi^2/\kappa}\rho.
\label{qndfgen}
\end{equation}
The second term on the r.h.s. of Eq.~(\ref{qndfgen}) 
is the 
usual double-commutator term associated to the 
measurement of 
$X$, it results from the elimination of the meter
variables; the third term is the feedback term itself and 
the fourth 
term is a diffusion-like term, which is an unavoidable 
consequence of the noise introduced 
by the feedback itself.

Due to the fact that the vibrational motion occurs at a
frequency $\nu $ of the order of MHz 
and the damping rate of the center-of-mass
motion, $\gamma$, is 
usually very small \cite{9},  it seems reasonable \cite{10} to 
use the generator ${\cal L}$ as that of
quantum optical master equation at a nonzero
temperature \cite{11}
(indicating the number of thermal phonons
with $n=[\exp(\hbar\nu/k_BT)-1]^{-1}$).
Moreover, 
since the Liouville superoperator ${\cal K}$ can only 
be of Hamiltonian 
form \cite{8}, it is taken as 
${\cal K}\rho =g 
\left[a-a^{\dagger},
\rho \right]/2$,
which means a driving 
term on the momentum, while
$g$ is the feedback gain related to the 
practical way of realizing the loop. 
Using the above expressions in
Eq.~(\ref{qndfgen}) and rearranging the terms in 
an appropriate 
way, it is possible to get the following master equation:
\begin{eqnarray}\label{totale}
&&\dot{\rho }=\frac{\Gamma}{2}(N+1)
\left(2a\rho a^{\dagger}-a^{\dagger}a\rho 
-\rho a^{\dagger}a\right)
+\frac{\Gamma}{2}N
\left(2a^{\dagger}\rho a-aa^{\dagger}\rho 
-\rho aa^{\dagger}\right)
 \\
&&-\frac{\Gamma}{2}M
\left(2a^{\dagger}\rho a^{\dagger}-a^{\dagger 2}\rho 
-\rho a^{\dagger 2}
\right)
-\frac{\Gamma}{2}M^{*}
\left(2a\rho a-a^{2}\rho -\rho a^{2}\right)
-\frac{g}{4}\sin\varphi
\left[a^{2}-a^{\dag 2},\rho\right]\,,\nonumber
\end{eqnarray}
where $\Gamma=\gamma-g\sin\varphi$, and
\begin{equation}\label{parameters}
N=\frac{1}{\Gamma }\left[\gamma n
+\frac{\chi^2 
}{4\kappa}+\frac{g^{2}}{4\eta\chi^2/\kappa}+\frac{g}{2}
\sin\varphi\right]\,;\;
M=-\frac{1}{\Gamma }\left[
\frac{\chi^2}{4\kappa}-\frac{g^{2}}{4\eta\chi^2/\kappa}
-i\frac{g}{2}\cos\varphi\right]\,.\nonumber
\end{equation}
Eq.~(\ref{totale}) clearly shows the 
effects of the feedback loop on the 
vibrational mode $a$. 
The proposed
feedback mechanism, indeed, 
not only introduces a driving term, but it also
simulates the presence of a bath with nonstandard 
fluctuations, characterized by an effective damping 
constant $\Gamma $ and by the coefficients
$M$ and $N$, which are given 
in terms of the feedback parameters \cite{6}. 
An interesting aspect of this effective bath is that 
it is characterized by 
phase-sensitive fluctuations.

Because of its linearity, the solution of  
Eq. (\ref{totale}) can be easily obtained
in terms of the normally ordered 
characteristic 
function ${\cal C}(\lambda,\lambda^*,t)$ \cite{11}.
The stationary solution has the following form
\begin{equation}\label{charsol}
{\cal C}(\lambda,\lambda^*,\infty)
=\exp\left[-\zeta|\lambda|^2+\frac{1}{2}\mu(\lambda^*)^2
+\frac{1}{2}\mu^*\lambda^2\right]\,,
\end{equation}
where
\begin{equation}\label{zemu}
\zeta=N+g\sin\varphi\frac{Ng\sin\varphi+\Gamma {\rm Re}\{M\}
+(g\sin\varphi)/2}
{\Gamma^2-g^2\sin^2\varphi}\,;\quad
\mu=\frac{\Gamma}{g\sin\varphi}\left(\zeta-N\right)\,.
\end{equation}

Under the stability conditions and in the long time 
limit $(t\to\infty)$
the variance of the generic quadrature operator
$X_{\theta}=(a e^{i\theta}+a^{\dag}e^{-i\theta})/2$ becomes
\begin{equation}\label{varXth}
4\langle X^2_{\theta}\rangle=1+2\zeta+
2{\rm Re}\{\mu e^{2i\theta}\}\,.
\end{equation}
For the position quadrature ($\theta=0$), it can be simply 
written as
\begin{equation}\label{varXg}
4\langle X^2\rangle=1+2n_{eff}\,,
\quad
n_{eff}=\zeta+{\rm Re}\{\mu\}\,.
\end{equation}
In absence of feddback ($g=0$) we have $n_{eff}\equiv n$, 
otherwise
$n_{eff}$ can be smaller than $n$ 
(the choice $\varphi=-\pi/2$ turns out to be the best), 
providing a {\it stochastic} localisation in the
position quadrature, i.e. a {\it confinement}. Depending on 
the external parameters,
it can also be negative (but it is always $n_{eff} \ge -1/2$) 
accounting for  
the possibility of going beyond the standard quantum limit. This is 
a relevant result of the present feedback scheme since it is able to
reduce not only the thermal fluctuations but even the quantum ones.

\section{The Practical Implementation}

Let us now consider a trapped ion/atom in a Lamb-Dicke regime,
in the presence of a standing wave field, and let us study 
different ways to implement the above model \cite{12}.
 
A Hamiltonian resembling Eq.(\ref{Hgen}) can be obtained
by considering an atom located at the node and
in resonant condition for which \cite{13}
\begin{equation}\label{Hini}
H=\hbar\Delta\sigma_z
+\hbar\nu a^{\dag}a+\hbar\chi\frac{(\sigma_++\sigma_-)}{2}X\,,
\end{equation}
with $\sigma_z$, $\sigma_{\pm}$ the Pauli operators for two-level system, 
and $\Delta$ the-small-detuning between the atomic and the standing 
wave frequency. The coupling constant 
is given by the Rabi frequency times the Lamb-Dicke 
parameter.
Hence, in this case, the electronic degree of freedom plays the role of meter,
i.e. $\sigma_x\equiv{\cal O}$ and $\sigma_{\pm}\equiv{\cal O}^{\pm}$.
By exploiting the resonance fluorescence it could be possible to 
get the quantity
$\Sigma_{\varphi }=\left(\sigma_-e^{-i\varphi }
+\sigma_+e^{i\varphi }\right)/2$ through homodyne detection of 
the field 
scattered by the ion along a certain direction \cite{14}.
In fact, the detected field may be written in terms of the 
dipole moment operator 
for the transition
$|-\rangle\leftrightarrow|+\rangle$ as 
$E^{(+)}_s(t)=\sqrt{\eta\kappa}\sigma_-(t)$ \cite{11},
where $\eta$ is an overall quantum efficiency accounting 
for the detection efficiency and the fact that only 
a fraction of the fluorescent light is collected and 
superimposed with a mode-matched oscillator.

Alternatively, the off resonant situation can be also exploited.
The starting Hamiltonian in this case implies the quantisation of the
cavity field \cite{11}, and reads
\begin{equation}\label{H2}
H=\hbar\Delta\sigma_z
+\hbar\nu a^{\dag}a
+\hbar\frac{2\epsilon^2}{\Delta}\sigma_z\,b^{\dag}b\sin^2
\left(\,{\overline k}X+\phi\right)\,
\end{equation}
where $b$, $b^{\dag}$ are the boson operators of the radiation mode,
${\overline k}$ is the dimensionless wave vector,
$\epsilon$ is the dipole coupling constant,
and $\Delta$ the-large-detuning.

By considering the internal atomic degree in the ground state 
and $\phi=\pi/4$, the leading term of Eq. (\ref{H2}) is
\begin{equation}\label{H3}
H=\hbar\nu a^{\dag}a
-\hbar\frac{2\epsilon^2}{\Delta}b^{\dag}b\left(\,{\overline k}X+1/2\right)\,.
\end{equation}
After linearization around the steady state of the cavity mode, 
the dynamics will be governed by an effective Hamiltonian formally identical 
to that of Eq.(\ref{Hgen})
\begin{equation}\label{Heff1}
H=\hbar\left(
-\hbar\epsilon^2/\Delta\right)b^{\dag}b
+\hbar\nu a^{\dag}a
+\hbar\chi YX\,,
\end{equation}
where $Y=(b+b^{\dag})/2$,
and $\chi=-4\beta{\overline k}\epsilon^2/\Delta$, with
$\beta$ the stationary value for the radiation amplitude 
(assumed real for semplicity).
In this case the meter is represented by the cavity mode,
i.e. ${\cal O}\equiv Y$, and ${\cal O}^{+}\equiv a^{\dag}$, ${\cal O}^{-}\equiv a$.
The homodyne measurement of the light outgoing from the cavity allows to obtain 
the quantity $Y_{\varphi}=(ae^{-i\varphi}+a^{\dag}e^{i\varphi})/2$ as desired.

Up to this point it has not considered how particular feedback
Hamiltonians could be implemented. Of course it is
important to be able to realize a term in the feedback Hamiltonian
proportional to momentum. 
This is not so straightforward, but here
possible ways in which it might be achieved are suggested. 
If the exact location of
the trap is not an important consideration, then shifts in the position
(being strictly equivalent to a linear momentum term in the Hamiltonian),
are achieved simply by shifting all the position dependent terms in the
Hamiltonian, in particular the trapping potential. This is a shift in the
origin of the coordinates, and, being a virtual shift in the position,
produces a term in the dynamical equation for the position proportional
to the rate at which the trap is being shifted. When the experimental
arrangement is such that the distance covered by the particle during the
cooling is negligibly small compared to the trapping apparatus this may
prove to be a very effective way of implementing a feedback Hamiltonian
linear in momentum. Another method would be to apply a large impulse to
the particle so that during one feedback time-step the particle is move
the desired distance, and an equal and opposite impulse is then applied
to reset the momentum. 
On the other hand, the use of laser pulses could be useful as well, since
in accordance with the 
theory of laser cooling, the light exerts
on the ion/atom a force proportional to its momentum.

\section{Conclusion}

Summarizing a stochastic control of the vibrational motion 
for a trapped particle via QND-mediated feedback has been 
proposed.
In principle the model could be extended to the three 
dimensional case.
The main limitation of such type of feedback is that
it only acts on the measured (indirectly) variable,
and it takes places only through a driving term in the 
variable conjugate to that measured.

However, there are many ways in which the measurement signal may be fed
back to affect the system. In general, at a given time, any integral of
the measurement record up until that time may be used to alter the system
Hamiltonian and affect the dynamics.
Hence the above schemes could be improved.
To this end a promising technique seems to be  
the feedback via estimation \cite{15}.

\medskip

\noindent {\bf Acknowledgements}
The author acknowledges fruitful collaboration with
P. Tombesi and D. Vitali at University of
Camerino, Italy.

\end{document}